\documentclass[a4paper,12pt]{article}
\usepackage{graphicx} 

\begin{document}

\title{{\Large \textbf{Re-examination of the size distribution of firms}}} 
\author{Taisei Kaizoji\thanks{Division of Social Sciences, International Christian University, Tokyo, Japan. E-mail: kaizoji@icu.ac.jp, Home page: 
http://subsite.icu.ac.jp/people/kaizoji/.}, 
Hiroshi Iyetomi\thanks{Department of Physics, Niigata University, Niigata, Japan.} 
and Yuichi Ikeda\thanks{Hitachi Research Institute, Hitachi Ltd., Tokyo Japan.}}
\date{}
\maketitle

\begin{abstract}
In this paper we address the question of the size distribution of firms. To this aim, we use the Bloomberg database comprising multinational firms within the years 1995-2003, and analyze the data of the sales and the total assets of the separate financial statement of the Japanese and the US companies, and make a comparison of the size distributions between the Japanese companies and the US companies. 
We find that (i) the size distribution of the US firms is approximately log-normal, in agreement with Gibrat's observation \cite{Gibrat}, and in contrast (ii) the size distribution of the Japanese firms is clearly not log-normal, and the upper tail of the size distribution follows the Pareto law. It agree with the predictions of the Simon model \cite{Simon}.\par
\end{abstract}

\section{Introduction}
There is a long tradition in the studies on the size distribution of firms since Robert Gibrat\cite{Gibrat} presented the first model of the dynamics of firm size and industry structure. 
Gibrat postulated the following simplest stochastic process to explain the skew distribution of firm's size. Le us consider a stochastic process $ \{x(t)\} $ indexed by time $ t = 0, 1, ... $, where $ x(t)$ is the firm's size at time $t$. Let $ \varepsilon(t) $ be an identically and independently distributed random variable denoting the growth rate between period $ t-1$ and $t$. If growth is proportionate, then 
\begin{equation}
x(t) = (1 + \varepsilon(t)) x(t-1). 
\end{equation}
or for small intervals
\begin{equation}
ln x(t) - ln x(t-1) \approx \varepsilon(t). 
\end{equation}
where $ ln(1+\varepsilon(t)) \approx \varepsilon(t) $. 
As a result, it follows that 
$ ln x(T) = ln x(0) + \sum^T_{t=1} \varepsilon(t) $. 
Frm the central limit theorem, $ ln x(T) $ is asymptotically normal distributed, and hence $ x(T) $ is asymptotically lognormally distributed, provided the shocks are independently distributed and small. In other words, in line with Gibrat's proposition, a proportionate stochastic growth process leads to a lognormal distribution. There is considerable evidence that the size distribution of firms is log-normal[1-5]. \par
On the other hand, other empirical studies show that the size distribution of firms is approximated closely by the Pareto distribution [6-12]. Moreover, recent empirical studies [13-17] show that the survival function of firm size follows a power law with an exponent which is close to unity, the so-called the Zipf's law \cite{Zipf}.  With respect to the distribution of a firm's size, Fujiwara, et. al. \cite{Fujiwara} and Aoyama, et. al. \cite{Aoyama} propose a resolution of the puzzle and show that proportionate growth processes can generate Zipf's law@
\footnote{With respect to the size distribution of cities, this puzzle is also considered by Xavier Gabaix \cite{Gabaix}.}.
The purpose of this paper is to reinvestigate the statistical properties of the size distribution of firms using a comprehensive database of corporate finance. To this aim, we use the Bloomberg database comprising multinational firms within the years 1995-2003, and analyze the data of the sales and the total assets of the separate financial statement of the Japanese and the US companies, and make a comparison of the size distributions between the Japanese companies and the US companies. 
We find that (i) the size distribution of the US firms is approximately log-normal, in agreement with Gibrat's observation \cite{Gibrat}, and in contrast (ii) the size distribution of the Japanese firms is clearly not log-normal, and the upper tail of the size distribution follows the Pareto law. It agree with the predictions of the Simon model \cite{Simon}.\par

\section{Data Analysis}
\subsection{ Corporate financial data}
The database used is the corporate financial data of multinational firms provided by Bloomberg Ltd. In this paper we use the data of sales and total assets of annual data of 12 years from 1992 to 2003 of the separate financial statements. We analyze the data of the Japanese companies and of the U.S. companies. The number of companies collected by the database has changed every year as shown in Table 1. \par

\begin{table}
\begin{center}
\begin{tabular}{ccc} \hline
{\it   Year   }  & {\it   U.S.   }  & {\it   Japan   } \\ \hline
1995 &  8328 & 2218 \\
1996 &  9246 & 2419 \\
1997 & 10181 & 2593 \\
1998 & 10481 & 2787 \\
1999 & 10348 & 3088 \\
2000 &  9734 & 3465 \\
2001 &  9030 & 3635 \\
2002 &  8529 & 3711 \\
2003 &  7811 & 3714 \\ \hline
\end{tabular}
\end{center}
\caption{The number of companies in the U.S. and in the Japan} 
\end{table} 

Our aim is to determine if the size distribution of firms follows the log-normal distribution that is created from the stochastic process proposed by Gibrat \cite{Gibrat}. 
To this aim, we take the logarithm of sales and total assets, and standardize the data,  
\begin{equation}
Z = \frac{ln X - \mu}{\sigma}. 
\end{equation}
If the variable of the firm's size $ X $ has a log-normal distribution, 
then the standardized variable $ Z $ has a standard normal distribution, 
\begin{equation}
P(Z)=\frac{1}{\surd{2 \phi}} e^{-(Z)^2/2}.
\end{equation}
defined over the interval $ (-\infty, +\infty) $. 
We perform normality tests which determine if a sample of data of standardized variable $ z $ fits a standard normal distribution below. 
As a normality test we select the ones that are best known, the Kolmogorov-Smirnov test and the Chi-square test. These tests compares the cumulative distribution of the data with the expected cumulative normal distribution\footnote{The chi-square test is an alternative to the Kolmogorov-Smirnov test. The chi-square test can be applied to discrete distributions. The Kolmogorov-Smirnov test is restricted to continuous distributions.}.

\subsection{The size distribution of the US companies} 

\subsubsection{The sales}

The best way to evaluate how far your data are from the normal distribution is to look at a graph and see if the distribution deviates grossly from the standard normal distribution. First of all, we 
standardize the data of the sales of the U.S. companies. The standardized sales $ S $ is defined as 
\begin{equation}
 S = \frac{(ln s - \mu_s)}{\sigma_s}. 
\end{equation}
where $ s $ denotes the annual sales, and $ \mu_s $, the mean of the logarithm of $ s $, and $ \sigma_s $, the variance of the logarithm of $ s $. Figure 1(a) shows the probability density function of the standardized sales $ S $ of the US companies for each of years in the 1995-2003 period. The probability density function is defined by $ P(S) $. If the distribution of the sales $ s $ is log-normal, then the distribution of $ S $ is normal. The solid line shows a standard normal distribution. We can tell at a glance that the lower tail of the distribution of the standardized sales $ S $ is long, but the upper tail of the distribution of $ S $ is short relative to normal. Figure 1(b) shows the survival function of the standardized sales $ S $ of the U.S. companies defined by $ P(S > x)$. The figure also show that the upper tail of the distribution of the standardized sales $ S $ is short compared with the normal distribution. \par

Next, we make certain of this point using the normal probability plot that is a graphical technique for assessing whether or not a data set is approximately normally distributed. Figure 2 shows the normal probability plot for the standardized sales $ S $ of the 10481 US companies in 1998. The data are plotted against a theoretical normal distribution in such a way that the points should form an approximate straight line. Departures from this straight line indicate departures from normality. 
The first few points show increasing departure from the fitted line below the line and the last few points show an increasing departure from the fitted line below the line. It means that the tail of the distribution of the standardized sales $ S $ is long in the lower tail and is short in the upper-tail relative to the normal distribution. This agrees with the empirical result by \cite{Stanley}\footnote{Stanley, et. al. \cite{Stanley} investigated the size distribution of firms using the data of the sales of 4071 North American manufacturing firms (SCI codes 2000-3999) on Compustat. They find that the upper tail of the size distribution of firms is too thin relative to the log normal rather than too fat. }.
We have done the same analysis for each year of the 9 year period from 1995 to 2003, and obtained similar results. Thus, we can reasonably conclude that the normal distribution does not provide an adequate fit for this data set of the standardized sales $ S $. \par

Finally we perform the two statistical tests for normality, the Kolmogorov-Smirnov test and the Chi-square test. In Table 2 the test statistics are summarized along with the kurtosis and the skewness of the distribution of the standardized sales. 
Note that the KS test is the Kolmogorov-Smyrnov statistic for the null hypothesis of normality and the CS test is the Chi-square test for the null hypothesis of normality; the figures in parentheses are the p-value of the test which denotes the smallest significance level at which a null hypothesis of normality can be rejected. The null hypothesis, that the population distribution is normal, is rejected for large values of both of the test statistics. \par
It is known that for a normal distribution, the population kurtosis is 3 and skewness is 0, the distribution of the standardized sales $ S $ are skewed to the left because the skewness is negative. In all cases, the p-value of the test statistics is equal to zero, so that the tests reject the null hypothesis of normality. \par

\begin{table}
\begin{center}
\begin{tabular}{ccccc} \hline
{\it Year}  & {\it kurtosis}  & {\it skewness}  & {\it KS test} & {\it CS test}\\ \hline
1995 & 3.61 &   -0.23 & 0.027 (0.0) & 240.53 (0.0) \\
1996 & 3.72 & 	-0.25 & 0.031 (0.0) & 290.61 (0.0)\\
1997 & 3.66  & 	-0.31 & 0.032 (0.0) & 325.67 (0.0)\\
1998 & 3.55 &   -0.32 & 0.029 (0.0) & 286.91 (0.0)\\
1999 & 3.56 & 	-0.34 & 0.030 (0.0) & 321.59 (0.0)\\
2000 & 3.63 & 	-0.38 & 0.032 (0.0) & 380.13 (0.0)\\
2001 & 3.71 &	-0.42 & 0.034 (0.0) & 330.84 (0.0)\\
2002 & 3.78 &	-0.43 & 0.035 (0.0) & 384.68 (0.0)\\
2003 & 3.68  &	-0.41 & 0.031 (0.0) & 324.90 (0.0)\\ \hline
\end{tabular}
\end{center}
\caption{Summary statistics on the standardized sales $ S $ of the U.S. companies}  
\end{table}

In brief, the sales distribution of the U.S. firms is not log-normal, and is skewed to the left relative to a log-normal distribution, and the upper tail is short and the lower tail is long relative to the tail of a log-normal distribution. 

\subsubsection{The total assets}
Figure 3(a) and Figure 3(b) show the probability density function and the survival function of the standardized total-assets $ A $. The standardized total assets $ A $ is defined as 
\begin{equation}
 A = \frac{(ln a - \mu_a)}{\sigma_a}. 
\end{equation}
where $ a $ denotes the total assets, and $ \mu_a $, the mean of the logarithm of $ a $, and $ \sigma_a $, the variance of the logarithm of $ a $. The solid line shows a standard normal distribution. The figures seem to show that both of the lower tail and the upper tail of the distribution of the standardized sales $ A $ is slightly fat compared with the normal distribution. 
Figure 4 shows the normal probability plot for the standardized total assets $ A $ in 1998. Visually, the probability plot shows a strongly linear pattern.  The fact that the points in the lower and upper extremes of the plot do not deviate significantly from the straight-line pattern indicates that there are not any significant outliers relative to a normal distribution. We perform the same analysis for each of the years in the 1995-2003 period and obtain a similar result. The plot demonstrates that the distribution of the standardized total assets $ A $ is tolerably close to the standard normal distribution.\par 
These are verified by the results of the tests for normality that are summarized in Table 3. We should note that the Kolmogorov-Smirnov test accepts the normality hypothesis for the 1997 data and the 1998 data at the 0.1 significance level. 

\begin{table}
\begin{center}
\begin{tabular}{ccccc} \hline
{\it Year}  & {\it kurtosis}  & {\it skewness}  & {\it KS test} & {\it CS test}\\ \hline

1995 & 3.28 & 0.09 & 0.014 (0.0628) & 176.59 (0.0) \\
1996 & 3.29 & 0.18 & 0.015 (0.0352) & 173.06 (0.0) \\
1997 & 3.37 & 0.09 & 0.011 (0.1896) & 145.96 (0.0002) \\
1998 & 3.24 & 0.03 & 0.009 (0.3416) & 143.43 (0.0004) \\
1999 & 3.46 &-0.07 & 0.012 (0.0861) & 123.69 (0.0129) \\
2000 & 3.55 &-0.19 & 0.024 (0.0)    & 189.95 (0.0) \\
2001 & 3.71 &-0.33 & 0.032 (0.0)    & 290.33 (0.0) \\
2002 & 3.82 &-0.38 & 0.034 (0.0)    & 358.44 (0.0) \\
2003 & 3.8  &-0.37 & 0.039 (0.0)    & 284.05 (0.0) \\ \hline
\end{tabular}
\end{center}
\caption{Summary statistics on the standardized total-assets $ A $ of the US companies}  
\end{table}
In brief, we cannot reject the idea that the distribution of the total assets $ A $ for the US companies has a log-normal distribution. 

\subsection{The size distribution of the Japanese companies} 
In this subsection we investigate the shape of the distributions of the sales and total assets for the Japanese companies. 

\subsubsection{The sales}
Figure 5(a) and Figure 5(b) show the probability density function and the survival function of the standardized sales $ S $ for the Japanese companies for the years in the 1995-2003 period. The solid line shows a standard normal distribution. We can tell at a glance that the upper tail of the distribution of $ S $ is long relative to normal. \par
Figure 6 shows the normal probability plot for the 1998 sales. The actual plot lies below the theoretical plot for the lower tail, and the actual plot lies above the theoretical plot for the upper tail. This means that the 1998 data of the standardized sales $ S $ of the Japanese companies has long tails relative to the normal distribution. We perform the same analysis for each of the years in the 1995-2003 period and obtain a similar result. \par

\begin{table}
\begin{center}
\begin{tabular}{ccccc} \hline
{\it Year}  & {\it kurtosis}  & {\it skewness}  & {\it KS test} & {\it CS test} \\ \hline
1995 & 3.42   &	0.58 & 0.05 (0.0) & 158.32 (0.0) \\
1996 & 3.82   &	0.47&  0.05 (0.0) & 178.14 (0.0) \\
1997 & 3.83   &	0.41 & 0.05 (0.0) & 172.35 (0.0) \\
1998 & 3.94   &	0.35 & 0.05 (0.0) & 178.54 (0.0) \\
1999 & 4.16   &	0.21 & 0.04 (0.0) & 164.73 (0.0) \\
2000 & 4.86   &	0.05 & 0.04 (0.0) & 178.57 (0.0) \\
2001 & 4.72   &	0.008 & 0.04 (0.0001) & 162.92 (0.0) \\
2002 & 3.66   & 0.25 &  0.04 (0.0001) & 173.77 (0.0) \\
2003 & 3.38   &	0.34 & 0.035 (0.0003) & 160.1 (0.0) \\ \hline
\end{tabular}
\end{center}
\caption{Summary statistics on the standardized sales $ S $ of the Japanese companies}  
\end{table}

The statistics are summarized in Table 4. Since the skewness is positive, the distribution of the standardized sales $ S $ is skewed to the right relative to normal distribution. This is the opposite of the result for the U.S. companies. In all cases, the p-value of the test statistics of the Kolmogorov-Smirnov test and the Chi-square test are equal to zero. Thus, the tests clearly reject the null hypothesis of normality of the distributions of $ S $.\par 
We can conclude that the tails, particularly the upper tail, of the sales distribution of the Japanese companies is long relative to the log-normal distribution.  

\subsubsection{The total assets}
Figure 7(a) and Figure 7(b) show the probability density function and the survival function of the standardized total asset $ A $ of the Japanese companies for each of years in the 1995-2003 period. We can tell at a glance that the upper tail of the distribution of $ A $ is apparently fat relative to normal. \par
Figure 8 shows the normal probability plot for the 1998 total assets. The normal probability plot indicates that the first few points show increasing departure from the fitted line above the line and last the few points show increasing departure from the fitted line above the line. This means that for the 1998 data of the standardized total asset of the Japanese companies the lower tail of the distribution is short and the lower tail is longer relative to the normal distribution. We perform the same analysis for each of the years in the 1995-2003 period. We see that the upper tail of the distribution of the standardized total-assets $ A $ is fat relative to the normal.\par

\begin{table}
\begin{center}
\begin{tabular}{cccccc} \hline
{\it Year} & {\it kurtosis} & {\it skewness}  & {\it KS test} & {\it CS test} & {\it power-law} \\ \hline
1995 & 4.24 &0.94 & 0.069 (0.0) & 275.58 (0.0) & 1.4E\\
1996 & 4.38 &0.90 & 0.072 (0.0) & 297.25 (0.0) & 1.7 \\
1997 & 4.38 &0.95 & 0.072 (0.0) & 325.35 (0.0) & 1.4 \\
1998 &4.43 &0.93 & 0.074 (0.0) & 337.1  (0.0) & 1.4 \\
1999 &3.96 &0.77 & 0.082 (0.0) & 475.89 (0.0) & 2.1 \\
2000 &3.98 &0.71 & 0.076 (0.0) & 462.35 (0.0) & 2.1 \\
2001 &4.03 &0.74 & 0.076 (0.0) & 463.72 (0.0) & 1.9\\
2002 &4.14 &0.71 & 0.073 (0.0) & 476.03 (0.0) & 1.8\\
2003 &4.02 &0.75 & 0.073 (0.0) & 497.95 (0.0) & 1.8\\ \hline
\end{tabular}
\end{center}
\caption{Summary statistics on the standardized total assets $ A $ of the Japanese companies}  
\end{table}
The statistics are summarized in Table 5. Since the skewness is positive, the distribution of the standardized total asset $ A $ is skewed to the right relative to normal distribution. In all cases, the p-value of the test statistics of the Kolmogorov-Smirnov test and the Chi-square test are equal to zero. Thus, the tests clearly reject the null hypothesis of normality of the distributions. We can conclude that the upper tail of the total-asset distribution of the Japanese companies is apparently long relative to a log-normal distribution, and the upper tail seems to follow a power law $ P(A > x) \sim x^{-\alpha} $ where $ \alpha $ denotes the power- law exponent. The last row of Table 4 shows the values of the power-law exponent $ \alpha $ which is estimated by the least square method. 

\section{Conclusion}
In this paper we investigate the size distribution of firm. 
In particular, we compare the size distributions of firms in Japan and in the U.S. with a log-normal distribution. In summary, we find the size distributions of the Japanese firms are not log-normal, and the total-assets distributions seem to follow a power-law in the upper tail. On the other hand, the size distribution of the US firms is well approximated by a log-normal distribution. Our findings make it clear that there is no universality of the size distribution of firms. The question is why the shape of the size distribution of the Japanese firms is different from those of the US firms. This calls for further consideration. 

\section{Acknowledgement} 
The authors wish to thank Prof. Hideaki Aoyama, Dr. Yoshi Fujiwara, and Dr. Wataru Soma for making a number of helpful suggestions. This research was supported by a grant from Hitachi Ltd. and the Japan Society for the Promotion of Science.

\begin{figure}
\begin{center}
  \includegraphics[height=14cm,width=12cm]{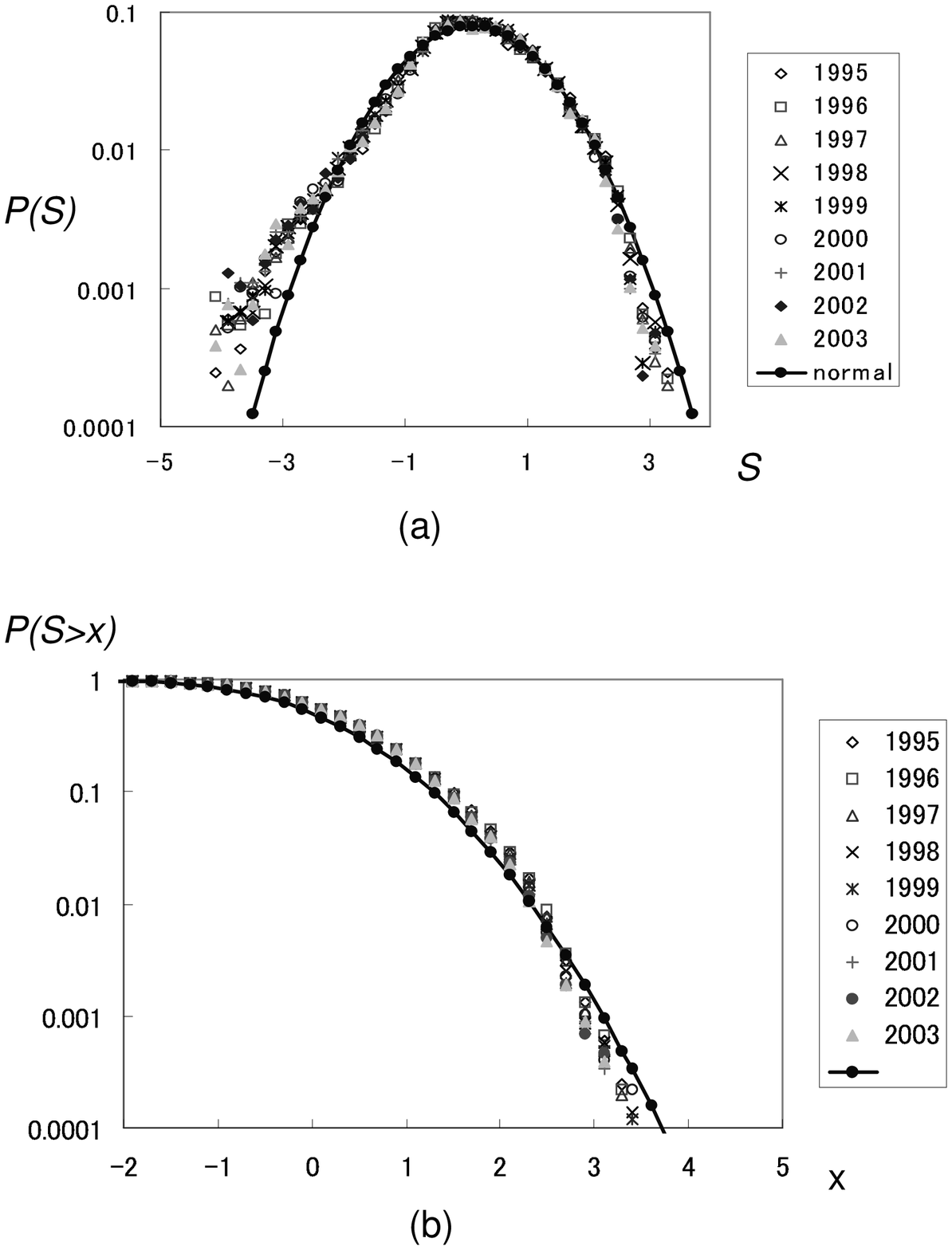}
\end{center}
\caption{(a) Probability density function and (b) the survival function of the logarithm of the standardized sales $ S $ for the US companies for each of the years in the 1995-2003 period. The standardized sales are defined by $ S = (ln s - \mu_s)/\sigma_s $ where $ s $ denotes the annual sales, and $ \mu_s $, the mean of the logarithm of $ s $, and $ \sigma_s $, the variance of the logarithm of $ s $. The solid lines show a standard normal distribution. }
\label{fig1}
\end{figure}
\newpage
\begin{figure}
\begin{center}
  \includegraphics[height=14cm,width=12cm]{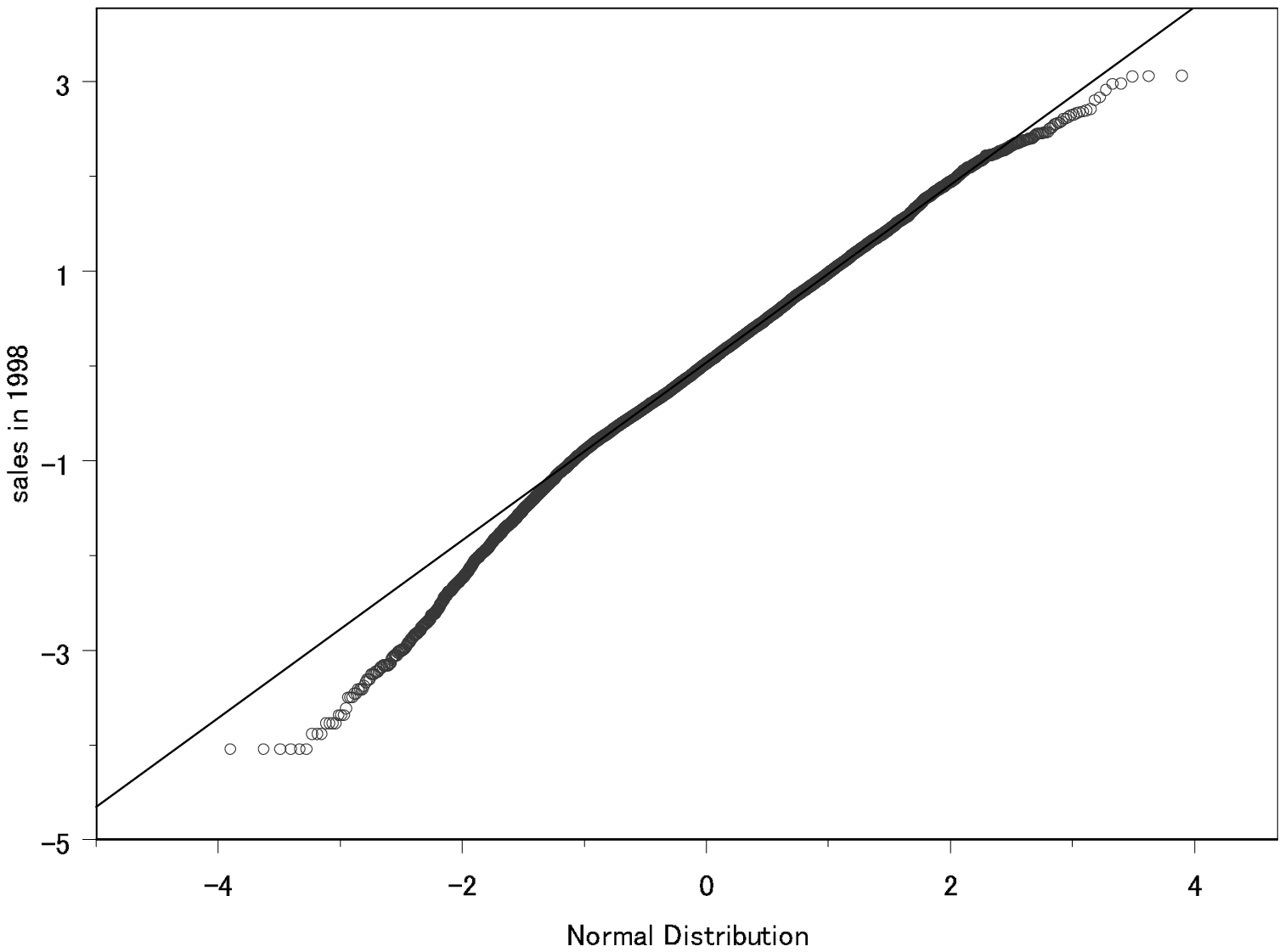}
\end{center}
\caption{The normal probability plot of the logarithm of the standardized sales $ S $ for the US companies in 1998. The standardized sales are defined by $ S = (ln s - \mu_s)/\sigma_s $ where $ s $ denotes the annual sales, and $ \mu_s $, the mean of the logarithm of $ s $, and $ \sigma_s $, the variance of the logarithm of $ s $. The straight line shows a standard normal distribution. }
\label{fig2}
\end{figure}
\newpage
\begin{figure}
\begin{center}
  \includegraphics[height=14cm,width=12cm]{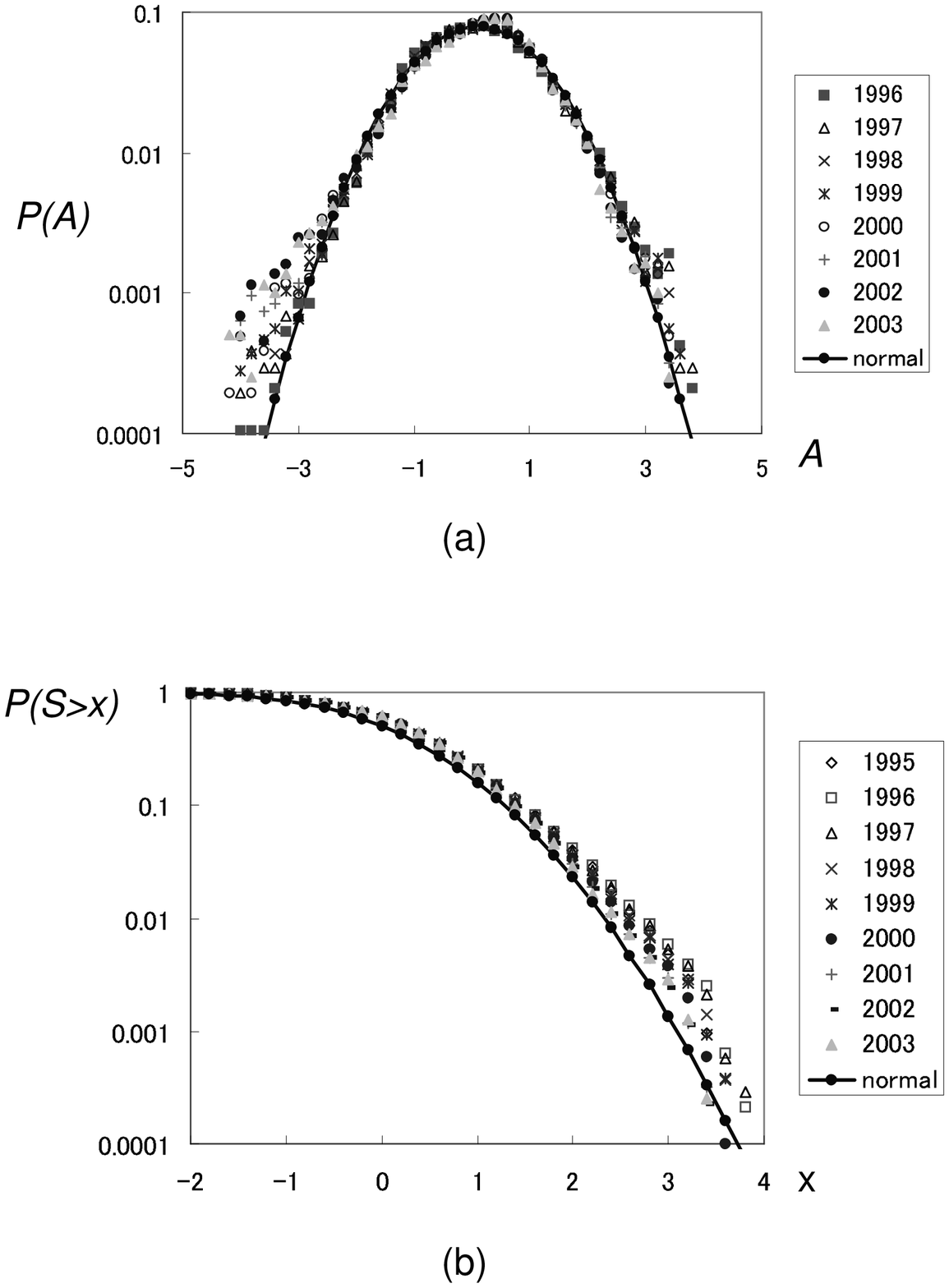}
\end{center}
\caption{(a) Probability density function and (b) the survival function of the standardized total-assets $ A $ for the US companies for each of the years in the 1995-2003 period. The standardized total-assets are defined by $ A = (ln s - \mu_a)/\sigma_a $ where $ a $ denotes the total assets, and $ \mu_a $, the mean of the logarithm of $ a $, and $ \sigma_a $, the variance of the logarithm of $ a $. The solid lines show a standard normal distribution. }
\label{fig3}
\end{figure}
\newpage
\begin{figure}
\begin{center}
\includegraphics[height=14cm,width=12cm]{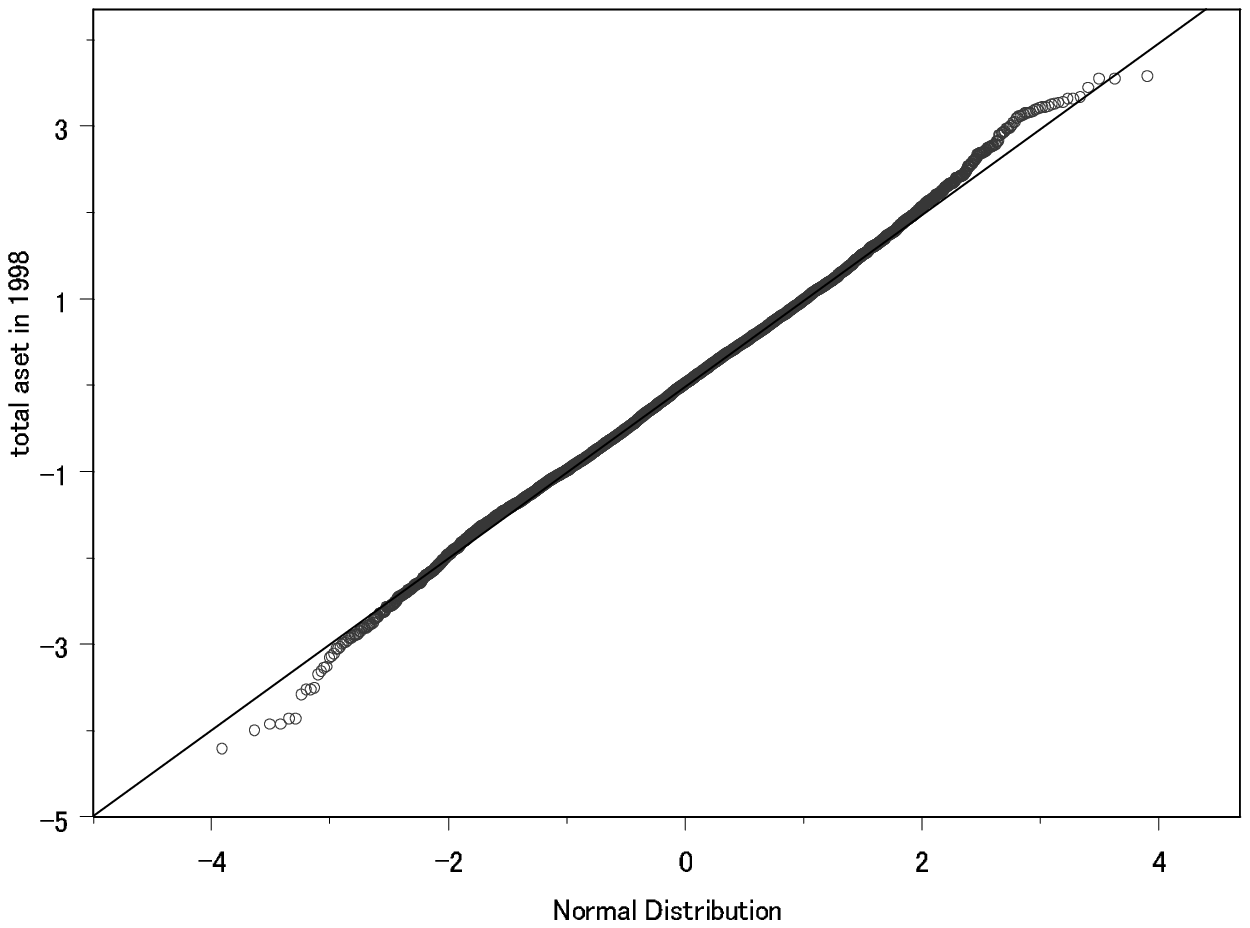}
\end{center}
\caption{The normal probability plot of the standardized total-assets $ A $ for the US companies in 1998. The standardized total-assets are defined by $ A = (ln s - \mu_a)/\sigma_a $ where $ a $ denotes the total assets, and $ \mu_a $, the mean of the logarithm of $ a $, and $ \sigma_a $, the variance of the logarithm of $ a $. The straight line shows a standard normal distribution. }
\label{fig4}
\end{figure}

\begin{figure}
\begin{center}
  \includegraphics[height=14cm,width=12cm]{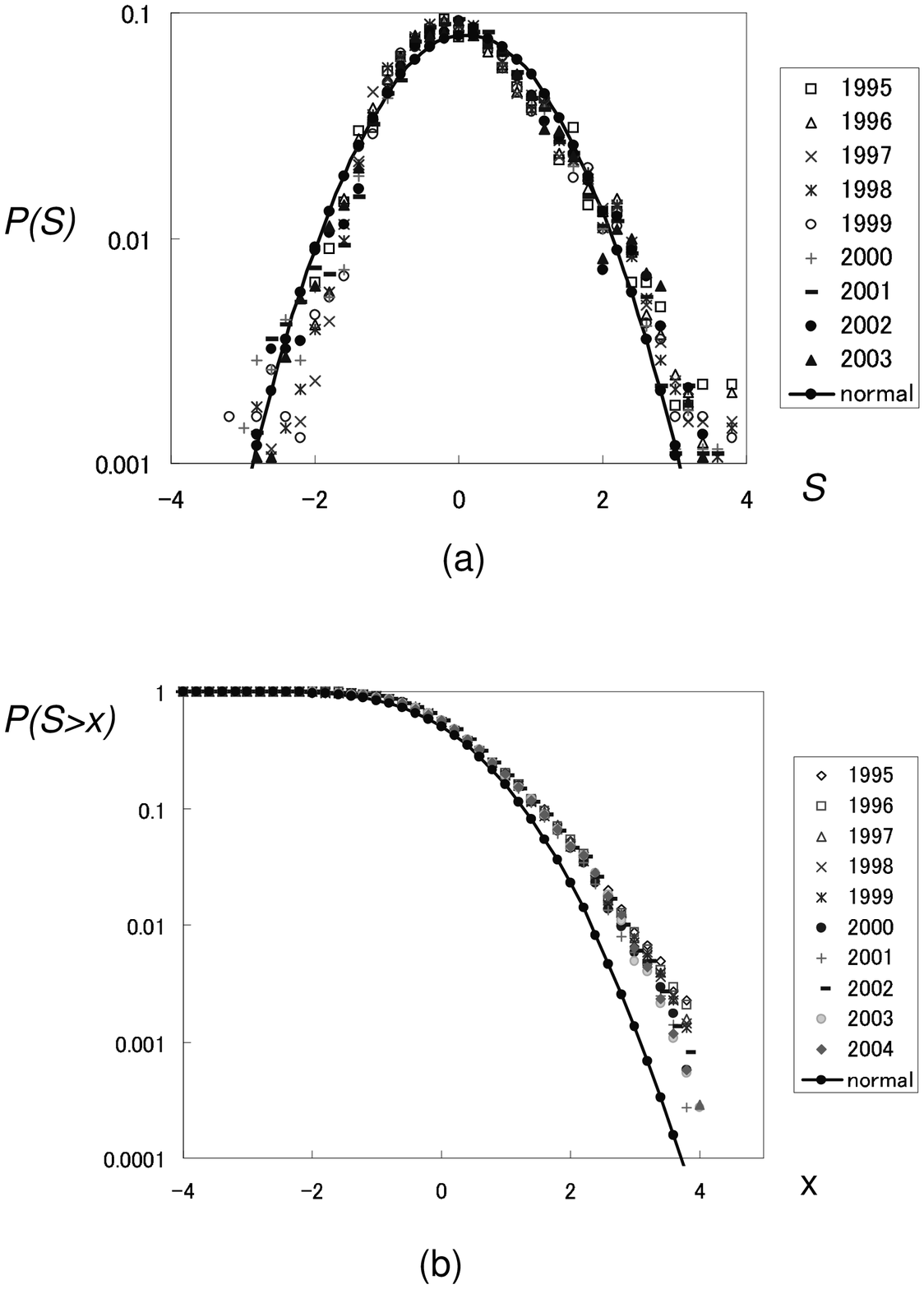}
\end{center}
\caption{(a) Probability density function and (b) the survival function of the standardized sales $ S $ for the Japanese companies for each of the years in the 1995-2003 period. The standardized sales are defined by $ S = (ln s - \mu_s)/\sigma_s $ where $ s $ denotes the annual sales, and $ \mu_s $, the mean of the logarithm of $ s $, and $ \sigma_s $, the variance of the logarithm of $ s $. The solid lines show a standard normal distribution. }
\label{fig5}
\end{figure}
\newpage
\begin{figure}
\begin{center}
  \includegraphics[height=14cm,width=12cm]{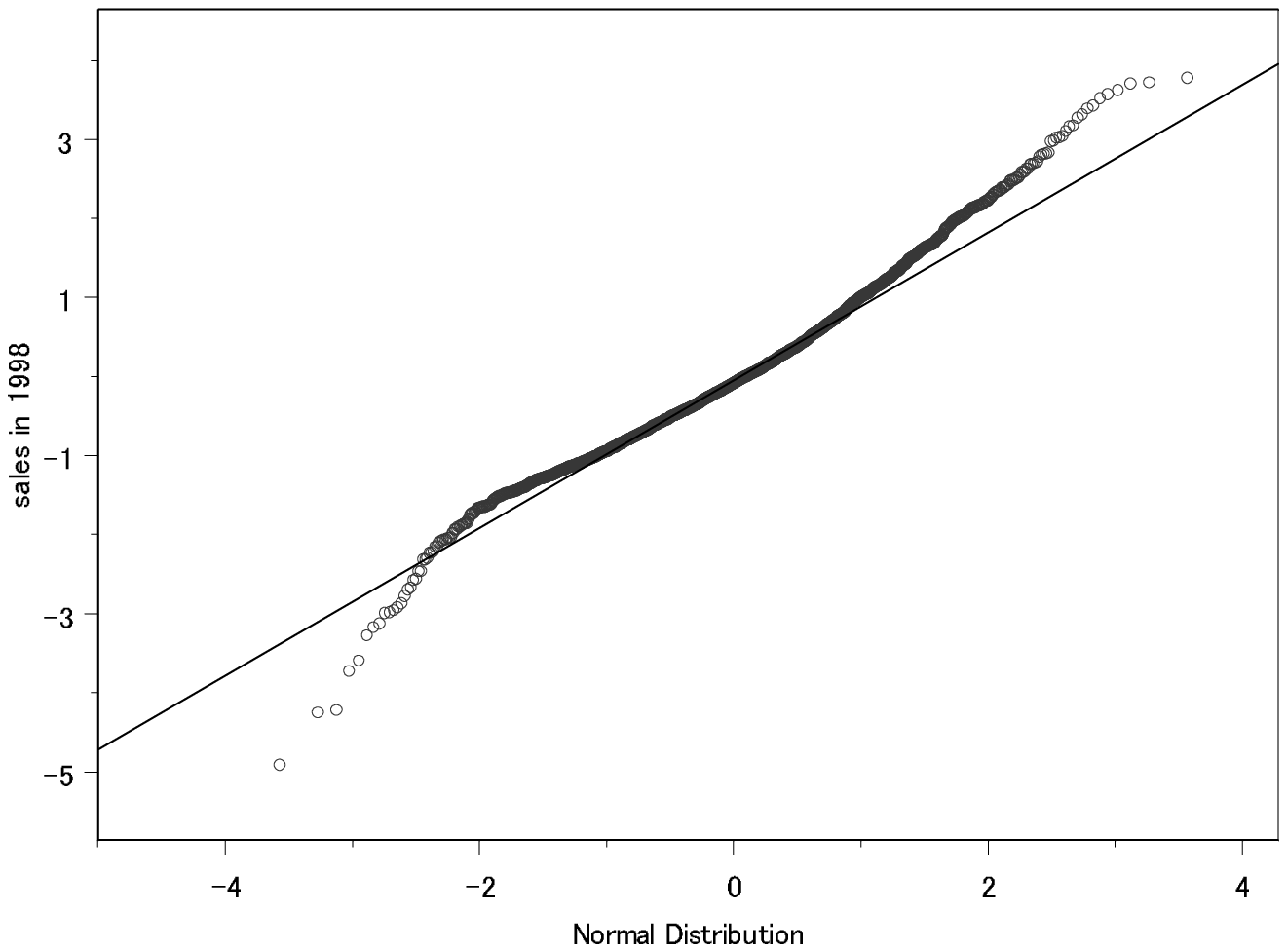}
\end{center}
\caption{The normal probability plot of the standardized sales $ S $ for the Japanese companies in 1998. The standardized sales are defined by $ S = (ln s - \mu_s)/\sigma_s $ where $ s $ denotes the annual sales, and $ \mu_s $, the mean of the logarithm of $ s $, and $ \sigma_s $, the variance of the logarithm of $ s $. The straight line shows a standard normal distribution. }
\label{fig6}
\end{figure}
\newpage
\begin{figure}
\begin{center}
  \includegraphics[height=14cm,width=12cm]{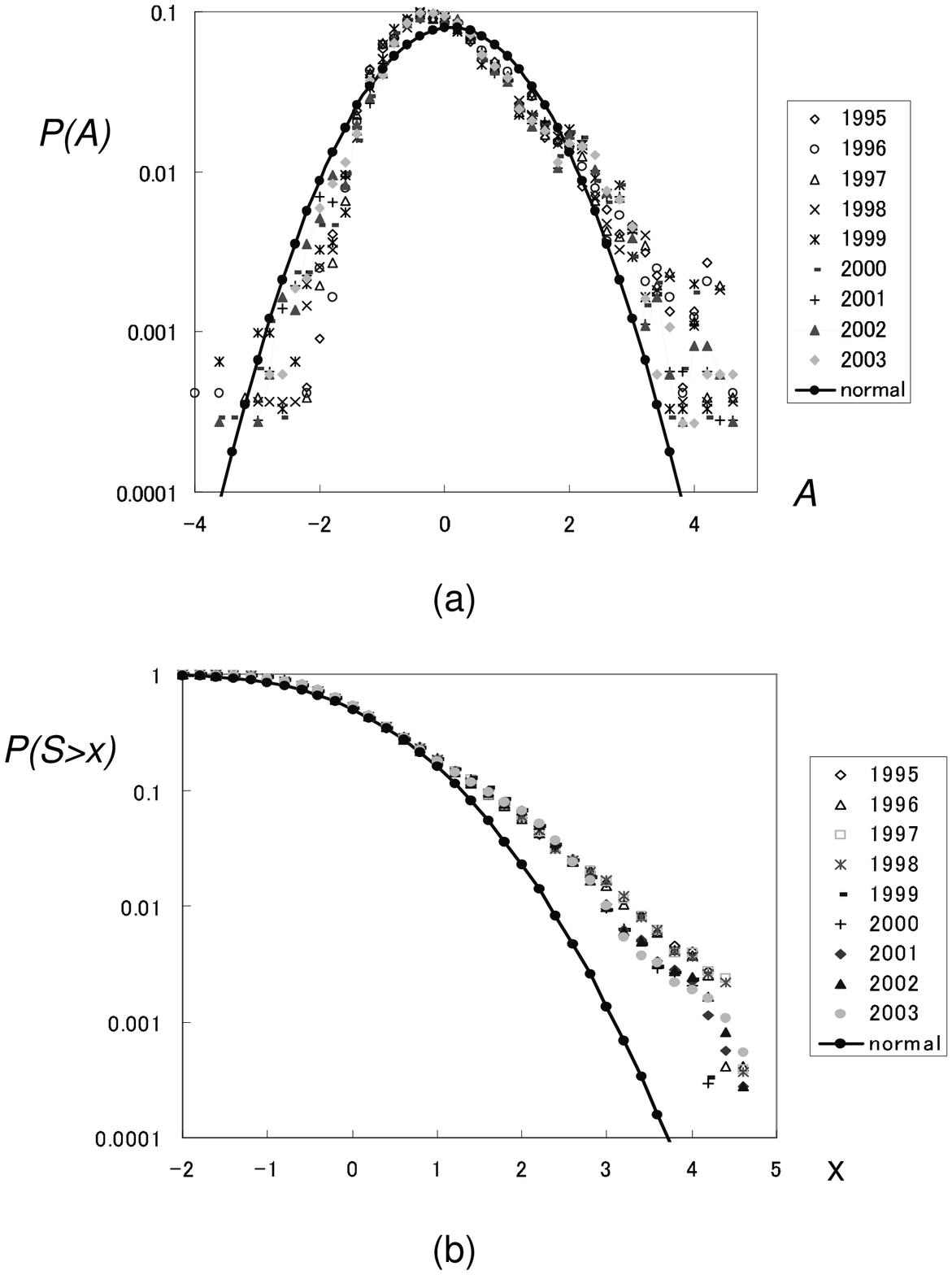}
\end{center}
\caption{(a) Probability density function and (b) the survival function of the logarithm of the standardized total-assets $ A $ for the Japanese companies for each of the years in the 1995-2003 period. The standardized total-assets are defined by $ A = (ln s - \mu_a)/\sigma_a $ where $ a $ denotes the total assets, and $ \mu_a $, the mean of the logarithm of $ a $, and $ \sigma_a $, the variance of the logarithm of $ a $. The solid lines show a standard normal distribution.}
\label{fig7}
\end{figure}
\newpage
\begin{figure}
\begin{center}
  \includegraphics[height=14cm,width=12cm]{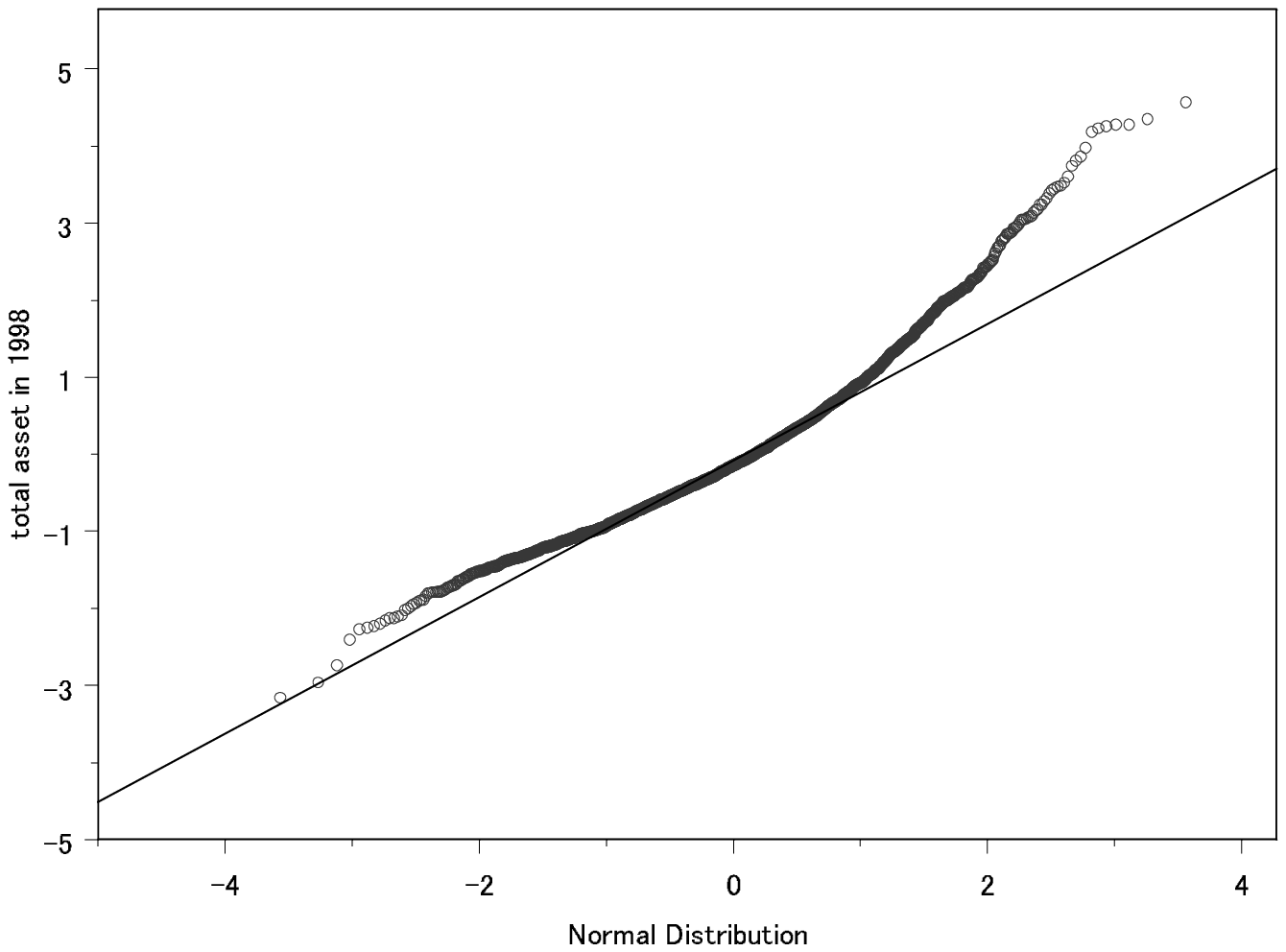}
\end{center}
\caption{The normal probability plot of the logarithm of the standardized total- assets $ A $ for the Japanese companies in 1998. The standardized total-assets are defined by $ A = (ln s - \mu_a)/\sigma_a $ where $ a $ denotes the total assets, and $ \mu_a $, the mean of the logarithm of $ a $, and $ \sigma_a $, the variance of the logarithm of $ a $. The straight line shows a standard normal distribution.}
\label{fig8}
\end{figure}
\end{document}